\documentstyle[preprint,aps]{revtex}
\begin{document}
\draft

%\begin{title}
\title{Localization-Delocalization Transition of Electron States in a
Disordered Quantum Small World Network}
%\end{title}
\author{Chen-Ping Zhu$^{1,2}$ and Shi-Jie Xiong $^1$}
%\begin{instit}
\address{ $^1$ National Laboratory of Solid State Microstructures and
Department of Physics,
Nanjing University,
Nanjing 210093, People's Republic of China\\
$^2$ College of Science, Nanjing University of Aeronautics and Astronautics,
Nanjing 210016,People's Republic of China}
%\end{instit}
\maketitle

\begin{abstract}
     We investigate the localization behavior of electrons
in a random lattice which is constructed from a
quasi-one-dimensional chain with large coordinate number $Z$ and
rewired bonds, resembling the small-world network proposed
recently but with site-energy disorder and quantum links instead
of classical ones.
 The random rewiring of bonds in the chain with large $Z$ enhances
both the topological disorder and the effective dimensionality.
From the competition between disorder and dimensionality
enhancement a transition from localization to delocalization is
found by using the level statistics method combined with the
finite-size scaling analysis. The critical value of the rewiring
rate for this transition is determined numerically. We obtain a
universal critical integrated distribution of level spacing $s$
in the form $I_{p_{c}}(s)\propto \exp (-A_{c}s^{\alpha})$, with
$A_{c}\simeq 1.50$ and $\alpha \simeq 1.0$. This reveals the
possible existence of metal-insulator transition in materials
with chains as the backbones.
\end{abstract}
\pacs{ PACS numbers: 73.61.Ph,05.60.+w,71.30.+h,72.90.+y }
\newpage

%\narrowtext
    \section{Introduction}

    Small-world network (SWN) model proposed by Watts and Strogatz \cite {s1}
has recently attracted much attention since it can describe
typical properties of diverse materials ranging from regular
lattices to random graphs simultaneously by varying only a single
parameter. Most of the previous works have focused on the average
minimal distance $\overline {l}$ separating two randomly chosen
points which describes how the distance is reduced by the
rewiring of the bonds in the network. Several numerical and
analytical studies have investigated the crossover behavior \cite
{s2}, the scaling properties \cite{s3}, and the percolation of
the dynamic processes \cite {s4} in this model. Although the
connection between any two vertices, no matter it is a regular
link or a ``shot-cut'', is treated as a classical one up till
now, the model has shown its wide applications on modeling
communication networks, disease spreading, and power grid, etc.
To stimulate out new applications of SWN in condensed matter
physics, it is interesting to introduce quantum connections
instead of classical ones among vertices and to investigate the
properties of electronic states. By rewiring the bonds from a
one-dimensional (1D) chain with large coordinate number, the
lattice becomes random and its effective dimension increases. If
the diagonal disorder is also included, it becomes a suitable
model to probe the localization and delocalization behavior for
realistic systems for which the backbones are 1D chains.

    The construction of a small-world network starts from a regular
1D chain with circular boundary condition, in which every site is
connected to its $Z$ immediate neighbors. Then every bond starting
from a site can be rewired with a probability $p$, i.e., to be
replaced by a new bond from the same site to a randomly chosen
site other than one of the $Z$ neighbors. From the geometric
point of view, the rewiring procedure introduces the topological
disorder to the originally regular 1D lattice but generates many
``shot-cut'' paths for moving of particles, causing easier
spreading of the states. At the same time the additional
site-energy disorder trends to localize the states. Therefore,
there are two competing effects on localization properties of
electronic states, one is the localization effect produced by the
topological and diagonal disorder, the other is the spreading
effect caused by the ``shot-cut'' links. In fact, such
``shot-cut'' links eventually lead to the enhancement of the
effective dimensionality of systems in the case of large $Z$.
Thus, a metal-insulator transition (MIT) can be expected by
varying $p$ if the effective dimensionality exceeds 2. The
purpose of this paper is to investigate localization properties
of electrons in such a disordered quantum small-world (DQSW)
model, and to seek for the possible transition between
localization and delocalization. We discern the localized and
delocalized states from the level statistics method combined with
the finite-size scaling analysis. We determine the transition and
obtain a universal critical distribution of level spacing at the
critical points, revealing the possible existence of MIT in some
soft materials.

    The paper is arranged as follows: in section II we describe the
    structure and basic
formalism of the DQSW model; in section III we show numerical
results on typical level spacing distributions; we present the
scaling analysis of the level statistics for the phase transition
in section IV; and we give a brief summary in the last section.

\section{Basic formalism}

    Consider an electron moving in a DQSW network,
the tight-binding  Hamiltonian of the system reads
\[
H=\sum_{i=1}^{N} \epsilon_{i}|i\rangle \langle i| +\sum_{i=1}^{N}\sum_{l=1}^{Z/2}
 [t_{1}(1-\rho_{i,l} )(|i\rangle \langle i+l|+|i+l\rangle \langle i|)
 \]
 \begin{equation}
 +t_{2}\rho_{i,l} (|i\rangle \langle i_R|+|i_R\rangle \langle i|)],
\end {equation}
where $\epsilon_i$ is energy level on site $i$, $N$ is the total
number of sites, $Z$ defines the immediate neighboring range
$[i-Z/2, i+Z/2]$ of size $i$, $i_R$ is an arbitrary site outside
this range and subject to the restriction that there is no
repeated term in the sum, and $t_1$ and $t_2$ are hopping
integrals for ``original" and ``rewired" links, respectively.
Here, $\epsilon_i$ and $\rho_{i,l}$ are random variables
satisfying the distribution probabilities
\begin{equation}
P(\epsilon_i )=\left\{ \begin{array}{l}
1/w ,\text{  if  } -w/2\leq \epsilon_i \leq  w/2,\\
0, \text{  otherwise},
\end{array}
\right.
\end{equation}
\begin{equation}
P(\rho_{i,l} )=\left\{ \begin{array}{l}
p, \text{  if  } \rho_{i,l}=1, \\
1-p, \text{  if  } \rho_{i,l}=0. \end{array} \right.
\end{equation}
Thus, $w$ characterizes the degree of on-site disorder as in the
Anderson model \cite {s5}and $p$ is the probability of a bond in
the immediate neighboring range to be broken and rewired. By
increasing $p$ the number of irregular bonds, which are also
``short-cut" paths for the motion of electrons, is increased,
leading to the enhancement of both the degree of topological
disorder and the effective dimensionality.

     Actually, the present model provides a way to
describe the random topological structures in some realistic soft
materials. In the case of polymer, the intrachain winding \cite
{s6} could serve as possible rewiring mechanism, and both
topological and site-energy disorder could exist. Moreover, the
circular boundary condition for a long chain which yields the
circle structure of a typical small-world network is also
applicable in analysis of the quasi-1D systems.

\section{Level statistics}

    The Hamiltonian matrix can be diagonalized numerically
    to yield single-electron
eigenvalue spectrum. It is well known that the levels of the
localized and extended states exhibit sharply different level
spacing distributions. The localized states have the exponential
decay of the space overlapping so that their levels are unrelated.
After unfolding \cite{s7,s8} the level spacing $s$ obeys Poisson
distribution
 $P_{\text{P}}(s)=\exp(-s)$ when the size of
system goes to infinity. Meanwhile the extended states of
consecutive energy levels have large space overlapping, resulting
in a correlated energy spectrum with level spacing satisfying
Wigner-Dyson (W-D) distribution $P_{\text{W-D}}(s)=\frac{\pi}{2} s
\exp(-\pi s^{2}/4)$.

 A DQSW system is characterized by the following parameters:
the degree of diagonal disorder $w$, the rewiring probability $p$,
the coordination number of the original 1D chain $Z$, the ratio
of the rewired hopping integral to the original one $t_2/t_1$, and
the size of system $N$. If $w\neq 0 $ and $p=0$, it is simply a 1D
disordered model with circular boundary condition and all the
states are localized. In this case $P(s)$ exhibits Poisson
distribution. By increasing $p$, the degree of the topological
disorder increases but the probability for an electron to tunnel
to farther sites is also enhanced, leading to the competition of
two opposite effects for the localization. The including of the
on-site disorder provides another mechanism of changing the
balance of the competition.
 In order to obtain a quantitative description of the resultant effect
of the competition, we adopt a scaling variable $\eta$ \cite {s9}
to depict the relative deviation of variance $\text{var}(s)$ of
level distribution $p(s)$ from that of the W-D distribution. The
variance is defined as
\begin {equation}
\text{var}(s) =<s^{2}>-<s>^{2},
\end{equation}
and the scaling function is
\begin{equation}
\eta (p,w,Z,N) =\frac {var(s)-0.273}{1-0.273},
\end{equation}
where $0.273$ and $1$ are the standard variance of W-D and
Poisson distributions, respectively. Thus $\eta $ can serve as a
measure of the deviation of states from the extended states. From
this definition we can calculate $p$ and $w$ dependence of $\eta$
for given values of other parameters. To suppress the fluctuations
in the results we take ensemble average over $10-100$ random
configurations. For the sake of simplicity we set $t_1$ as the
energy units. The degree of on-site disorder $w$ varies from $0$
to $22$, and the rewiring probability $p$ is ranged from $0$ to
$0.95$. In the calculations we take $Z=8$ or $4$ and $t_2=1$,
$0.3$, or $0.1$ to check their effect on the localization.
Meanwhile the scaling analysis is carried out by varying the
system size from $N=1200$ to $N=3600$.

In Fig. 1 we show the level spacing distribution of systems with
different rewiring probability and fixed values of $w$, $Z$, $t_2$
and $N$. By increasing the rewiring probability $p$ from zero
$P(s)$ varies from Poisson-like to Wigner-Dyson-like for a finite
$w$, reflecting the delocalization effect of the rewiring. It is
easily understood from the scaling theory \cite{a1} that the
states are localized for system with zero $p$ and finite $w$
because of its 1D nature. By increasing $p$ both the degree of
topological disorder and the effective dimensionality of the
system are enhanced. The results of Fig. 1 indicate that the
latter is dominant for $Z=8$, $t_2=1$ and the adopted range of
$p$. The appearance of the extended states is possible if the
effective dimensionality exceeds 2. Although Fig. 1 only shows
the behavior of system with fixed size, the typical W-D form of
the distributions for $p\sim 0.2$ suggests the extended nature of
the states. We will carry out the scaling analysis to check the
existence of MIT. Note that the crossing point $s_{0}\simeq 2.0$
of the curves is independent of the parameters, suggesting the
applicability of the level statistics for the present model.

The values of other parameters, such as $Z$ and $t_2$, also have
crucial effect on the localization behavior of the states.
Generally speaking, increasing the coordinate number $Z$ will
accelerate the tendency of delocalization in increasing $p$ from
zero. It is difficult to find the extended states if $Z\leq 4$.
The value of $t_2$, the hopping intensity of the rewired bonds,
has similar effect on the localization effect of the states. The
extended states can not be found if $t_2$ is too small. The degree
of topological disorder is enhanced by decreasing the ratio
$t_2/t_1$ from 1. Thus, in increasing $p$ the effect of enlarging
effective dimensionality may be cancelled by the increase of the
disorder. At this point the DQSW model is essentially different
from the classical SWN, in which the properties depend only on
the structural parameters. In Fig. 2 we plot the dependence of
scaling variable $\eta $ on parameters $w$ and $p$ for various
system sizes and other choices of $Z$ and $t_2$. It can be seen
from the main panel that $\eta$ varies monotonically with increase
of $w$ for all the investigated system sizes, reflecting the
trivial localization effect of the on-site disorder as in other
models. Although $\eta $ is certainly near zero for small $w$,
$\eta$ still decreases by increasing the system size, suggesting
the absence of the extended states in the thermodynamical limit.
This implies that for $Z=4$, $t_2=0.3$ and $p=0.04$, the effective
dimensionality could not exceed 2. From the inset it can be seen
that the effect of $p$ is rather nontrivial in the case of
$t_{2}=0.1$ and $Z=4$. Since the rewired links have much weaker
hopping strength than the original ones, the $p$ dependence of
$\eta$ is no longer monotonous. It first decreases with $p$ and
reaches a minimum value $\eta_{m}$ at a medium value $p_{m}$,
then increases and finally drop down again near $p=1$. This
behavior is not changed by changing the system size. When $p$ is
small ($p<p_{m}$), the increase of $p$ creates the long-range
hopping paths and causes the states to be more expanded. For large
$p$ the hopping strength of the rewired links becomes important
and by increasing $p$ more bonds with large hopping intensity are
converted to ones with small hopping intensity. This enhances the
localization trend and increases the value of $\eta $. When $p$
is near 1, the structure approaches to the limit of a random
graph in which the 1D backbone completely disappears and the
original links with $t_1=1$ are almost absent. In this case the
increase of long-range paths can result in the increase of $\eta$.
All the states are still localized as shown from the $N$
dependence of $\eta$. From comparison of the inset and the main
panel, the value of $\eta $ is increased by decreasing $t_2$ from
0.3 to 0.1. It should be stressed that the $p$ dependence of
$\eta $ is sensitive to the value of $t_2$. If $t_2 > 1$ $\eta $
is monotonically increased with $p$ because by increasing $p$
more bonds are converted from weak hopping to strong hopping.

\section{Scaling behavior of level spacing distribution}

     In a new version of small-world model proposed by Newman and Wattz
\cite{s4}, the length scale $\xi$ defined as $\xi=\frac{1}{p}$
for the 1D underlying lattice governs the features of a number of
quantities, such as the mean distance of pairs of vertices, the
effective dimension $D$, etc. In their model $D$ is a
size-dependent quantity and related to the length scale in the
form $D=\log (p ZN)$. In the quantum version of small-world
network, $p$ still plays an important role in determine the
localization properties of electrons, although we adopt the
structure of the earlier version of the model proposed in
\cite{s1} and include the diagonal disorder. One can conjecture
that by the random rewiring procedure the effective dimension is
still related to $Z$ and $p$, although the relation $D=\log
(pZN)$ may no longer be valid. In this sense for large $Z$ and
suitable $p$ the effective dimensionality of DQSW can exceed $2$.
One could predict the occurrence of MIT in such systems without
violating the scaling hypothesis.

In Fig. 3 we plot the $p$ dependence of $\eta$ for systems with
varying size ($N= 1200, 1600 ,2400 $ and $3600$) and with
parameters $t_1=t_2=1$, $Z=8$ and $w=22$. The curves
corresponding to different system sizes are crossed at point $p_c
\sim 0.085$ for which $\eta(p)$ is size-independent and equal to
$\eta_c=0.37$. $p_c$ can be regarded as the transition point
separating the localization regime ($p < p_c$) and delocalization
regime ($p > p_c$). We also find that this transition can occur
in a range of $w$. The $w$ dependence of $p_{c}$ in $z=8$ is
shown in the inset of Fig. 3. $p_c$ increases with $w$ as can be
expected from the trivial effect of $w$.

The values of $\eta $ shown in Fig. 3 as a function of $p$ and
$L$ can be fitted with a one-parameter scaling function
\begin{equation}
\eta (L,p)= f(L/\xi (p)),
\end{equation}
where $\xi (p)$ can be regarded as the localization length in the
localized regime and the correlation length in the delocalized
regime. We set up the value of $\xi (p)$ in such a way that all
the $\eta (L,p)$ - $L/\xi (p)$ curves from the data of Fig. 3 can
collapse in a common curve representing the function $f(L/\xi
(p))$. In the inset of Fig. 4 we plot this function. It can be
seen that the curve consists of two branches: the upper branch
corresponds to the localization regime ($p< p_c$) and the lower
branch stands for the delocalization regime ($p> p_c$). The value
of $\xi $ is singular at the transition point $p_c$. This
singularity can be expressed by a power law with exponent $\nu $
\begin{equation}
\xi (p) =\xi_0 |p-p_c|^{-\nu}
\end{equation}
where $\xi_0 $ is a constant. By fitting the data we obtain that
the exponent is equal to $\nu =0.92 \pm 0.15$.

It is interesting to investigate the level statistics at the
critical points. For this purpose it is more convenient to
consider the cumulative level spacing distribution function
defined in the form \cite{s9} $I(s)=\int_{s'}^{\infty}p(s')ds'$.
From this definition one has $I_{p}(s)=\exp(-s)$ and
$I_{w}(s)=\exp (-\pi s^{2}/4)$ for the Poisson and W-D
distributions, respectively. To demonstrate the universal feature
of the distribution at the critical points, in Fig. 4 we plot the
$-\ln I(s)$-$s$ curves for parameters in the localization and
delocalization regimes and at the critical points. We find that
in the range of $s\geq 0.5$ all the curves corresponding to the
critical points for various values of $N$ and $w$ coincide with a
common straight line which can fitted by
\begin {equation}
I_{c}(s) \propto \exp (-A_{c}s^{\alpha})
\end {equation}
with coefficient $A_{c}=1.50\pm 0.06$ and exponent $\alpha=1.0$
independent of the values of $w$ and $N$. Since $\alpha =1$ the
critical distribution $P_{\text{c}}(s)$ is similar to the Poisson
distribution in the tail of large $s$. This feature of the
critical distribution at large $s$ has been obtained in the
Anderson transition in system of very large size \cite{a2}. In
the range $s < 0.5$, the curves $-\ln I(s)$-$s$ for the critical
points is deviated from the straight line and approaches to a
function of the form of Eq. (\ref{dis}) but with $\alpha $
greater than 1. This is a behavior that interpolates between
poisson and W-D distributions. Such a universal form for the
critical points provides the further evidence for the existence
of the transition in the DQSW system. Moreover, it is a successful
practice of the new method using the level statistics combined
with the finite size scaling analysis proposed in Ref. \cite{s9}
to determine the localization-delocalization transition in
systems with complicated structures.

\section{Discussion and Conclusions}

    We have investigated the small-world model in the viewpoint of
quantum Hamiltonian with the primary topological disorder in the
network and the additional diagonal disorder. Similarly to the
case of the earlier works which focus on the classical behavior
of this structure, we find that the rewiring procedure with
probability $p$ not only introduces the topological disorder, but
also enlarges the effective dimensionality of the system. As a
result of the competition between these two effects, the system
undergoes a metal-insulator phase transition by varying $p$. By
using the method of level statistics combined with the finite
size scaling analysis, we have determined the $w$-dependent
transition point $p_c$ in the DQSW system. It corresponds to a
transition of the level distribution $P(s)$ from the Poisson-like
form to the Wigner-Dyson-like form. For this transition a
two-branch scaling function is obtained. The calculated critical
point $p_{c}$ increases with increasing the diagonal disorder
$w$. Moreover, it belongs to a universal sort of phase transition
characterized by the critical distribution $\ln I_{p_c}(s)
\propto -1.5s $ at large $s$. The existence of such a transition
depends crucially on values of other parameters such as $Z$ and
$t_2$. When the hopping intensity of rewired bonds $t_{2}$ is too
small, the disorder effect of the rewiring process becomes
dominant and no extended states can be found. Furthermore, the
transition usually does not occur if $Z\leq 4$, because in this
case the effective dimensionality can not exceed 2 in the rewiring
process.

\section*{Acknowledgments}
We would like to thank S.N. Evangelou for useful discussion. This
work was supported by National Foundation of Natural Science in
China Grant No. 69876020 and by the China State Key Projects of
Basic Research (G1999064509).

\begin{center}
  {\bf Figure Captions}
  \end{center}

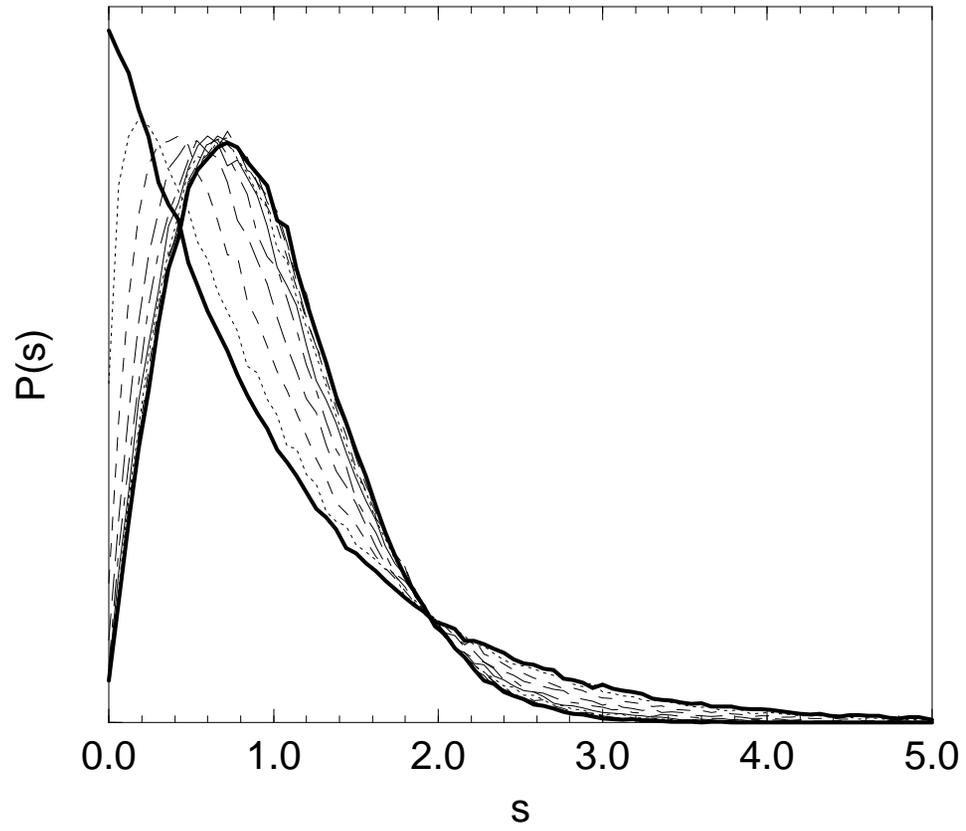
       {\bf Fig. 1} Level spacing distribution $P(s)$
for system with parameters $N=1600$, $w=16.0$, $Z=8$, and $t_2=1$.
$t_1$ is set to be the energy units. The rewiring probability
varies successively as $p=0$, $0.02$, $0.04$, $0.06$, $0.08$,
$0.10$, $0.12$, $0.14$, $0.16$, $0.18$, and $0.20$ for curves from
the Wigner-Dyson-like form to the Poisson-like form. The curves
are crossed at the common point $s_{0}\simeq 2.0$.

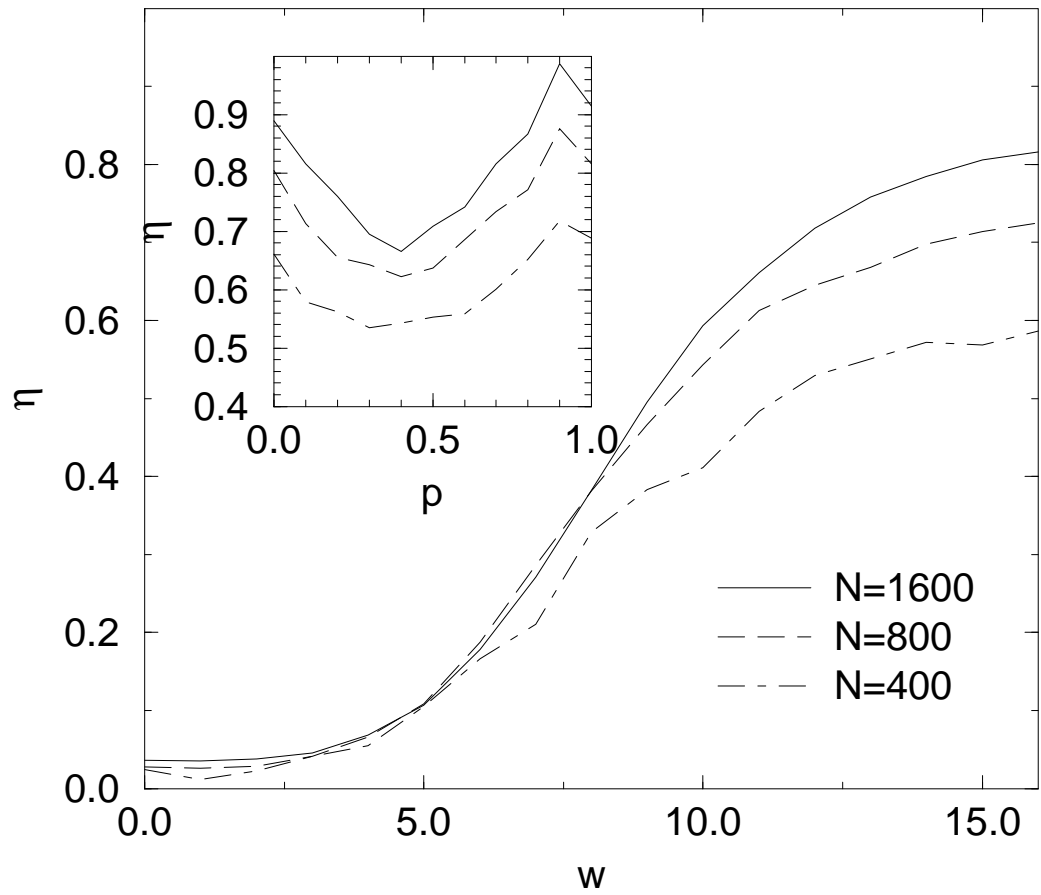
       {\bf Fig. 2} Scaling variable $\eta$ as a function of $w$ for
systems with $t_{2}=0.3$, $Z=4$, and $p=0.04$. The size of system
is $N=400$ (dot-dashed line), $N=800$ (long-dashed line), and
$N=1600$ (solid line). Inset: $\eta$ versus $p$ curves for
systems with $w=8$, $t_2=0.1$ and the same values of other
parameters as those in the main panel.

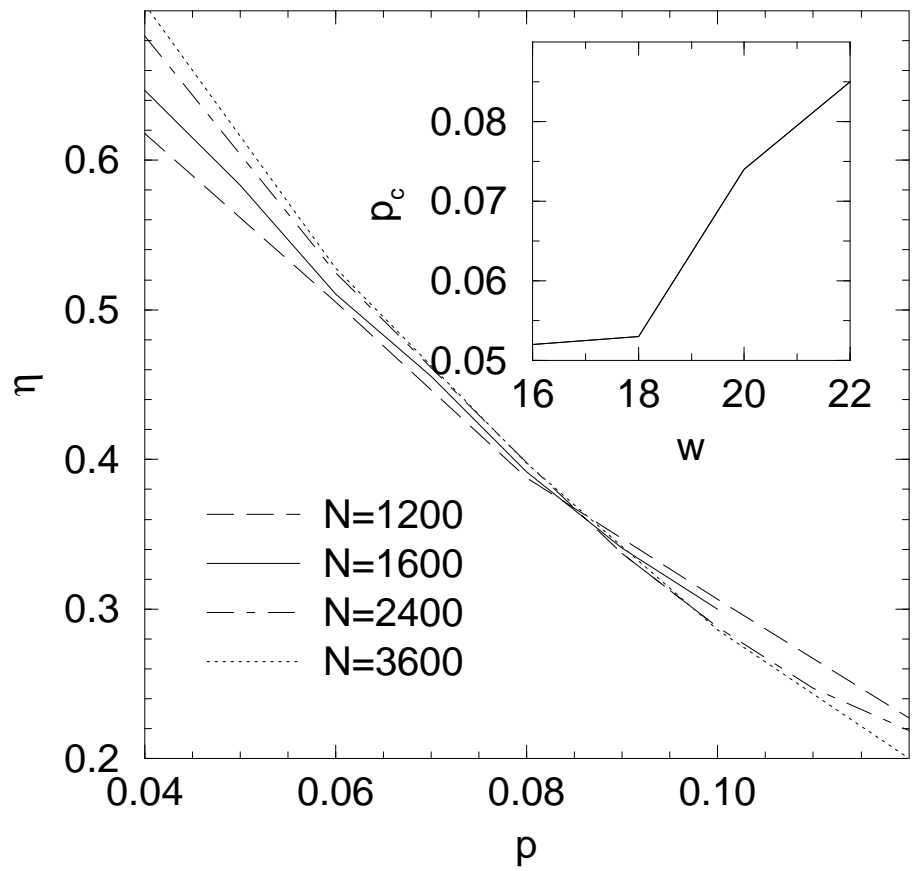
       {\bf Fig. 3} Scaling variable $\eta$ as a function of $p$ for
systems with $Z=8$, $w=22$, $t_{2}=1.0$ and various system sizes
$N$. Critical rewiring probability $p_{c}=0.085$ is determined
from the crossing point. Inset: Critical rewiring probability
$p_{c}$ as a function of the on-site disorder $w$.

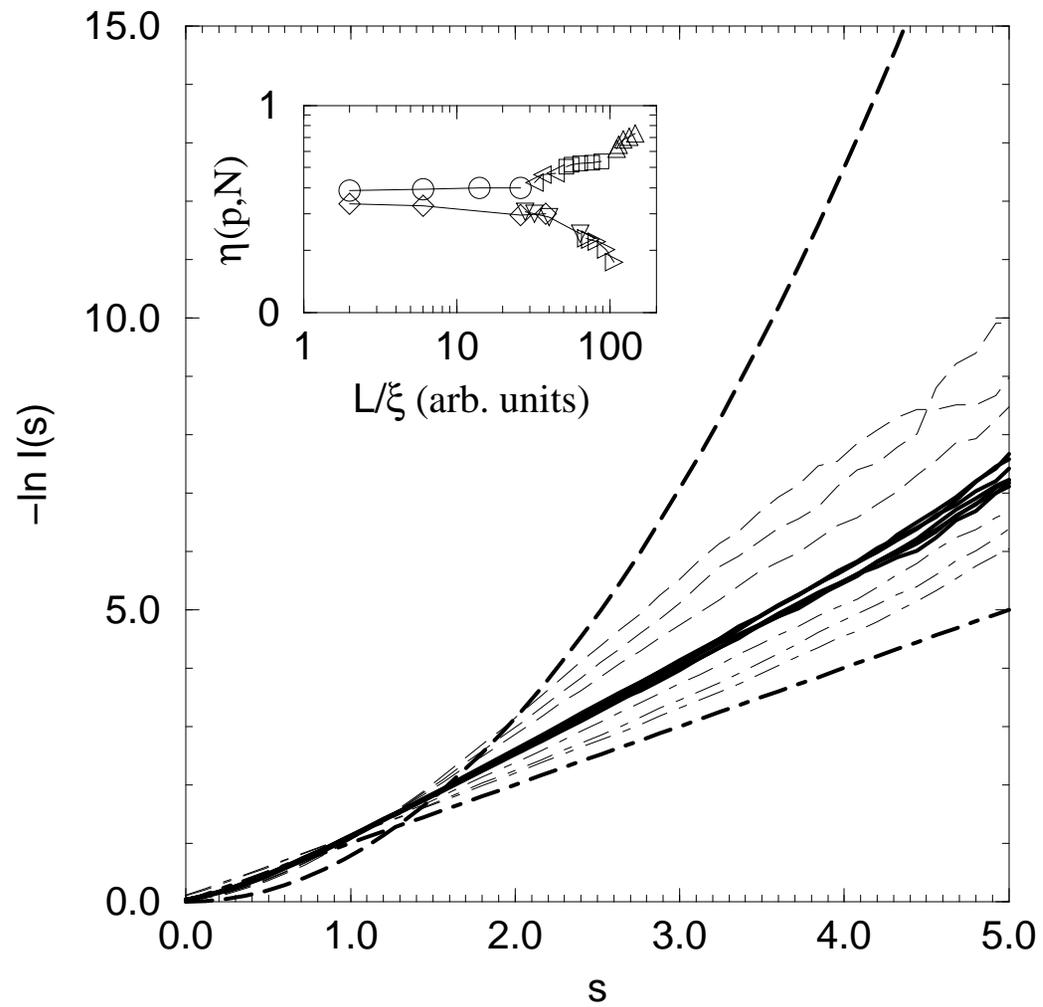
       {\bf Fig. 4}      Logarithmic integrated level spacing
distribution $-\ln (I_{c}(s))$ at the critical points $p_{c}$ for
systems with different $N$ and $w$ (thick solid curves). The
distributions for systems in the non-critical regimes are shown by
the thin dashed lines (delocalization regime) and the thin
dot-dashed lines (localization regime). Delocalization regime:
$N=1200$, $w=18$, $p=0.12$; $N=1600$, $w=18$, $p=0.10$; and
$N=1600$, $w=22$, $p=0.20$. Localization regime: $N=1600$, $w=18$,
$p=0.03$; $N=1200$, $w=22$, $p=0.02$; and $N=1600$, $w=22$,
$p=0.01$. The Poisson and Wigner-Dyson distributions are plotted
as references with thick dot-dashed line and thick dashed line,
respectively. All the distributions at the critical points for
$s>0.5$ collapse in a common straight line fitted by $-\ln
I_{c}(s) = 1.5 s -1.0 $. Inset: One-parameter scaling function
$\eta (L, p)$ versus $L/\xi $.

\end{document}